\documentclass[preprint,prb]{revtex4}
\usepackage{ulem,graphicx}

\begin{document}
\title{Direct generation of charge carriers in c-Si solar cells due to embedded nanoparticles}
\author{M. Kirkengen} \affiliation{Department of Physics, University
of Oslo,  P. O. Box 1048 Blindern, 0316 Oslo, Norway}
\author{J. Bergli} \affiliation{Department of Physics, University of
Oslo,  P. O. Box 1048 Blindern, 0316 Oslo, Norway} \author{Y. M.
Galperin} \affiliation{Department of Physics, University of Oslo,
P. O. Box 1048 Blindern, 0316 Oslo, Norway} \affiliation{Center for
Advanced Materials and Nanotechnology at the University of Oslo,
Argonne National Laboratory, 9700 S. Cass Ave., Argonne, IL 60439,
USA, and A. F. Ioffe Physico-Technical  Institute, 194021
St. Petersburg, Russia}
\date{\today}
\begin{abstract}
  It is known that silicon is an indirect band gap material, reducing
  its efficiency in photovoltaic applications. Using surface plasmons
  in metallic nanoparticles embedded in a solar cell has recently been
  proposed as a way to increase the efficiency of thin film silicon
  solar cells. The dipole mode that dominates the plasmons in small
  particles produces an electric field having Fourier components with
  all wave numbers. In this work, we show that such a field
  creates electron-hole-pairs without phonon assistance, and discuss
  the importance of this effect compared to radiation from the
  particle and losses due to heating.
\end{abstract}
 \maketitle
\section{Introduction}

Present day solar cell industry is completely dominated by the use of
silicon as the active material.   It's main advantages are
availability, disposability and several decades of industrial
metallurgical development, compared to the poisonous or rare elements
of, e.g., GaAs.  However, silicon is not an ideal material for solar
cells. One disadvantage  is that it has an indirect band gap. This
means that photons with energy  close to the band gap can only be
absorbed in phonon-assisted processes.  Therefore, the absorption of
these photons is weak, and the silicon wafer can not be made too thin
if one is to absorb this part of the solar spectrum.  The material
costs and limited production capacity for solar grade silicon mean
that the thickness required in todays first generation solar cells is
a significant obstacle to their commercial  success. Also, due to the
limited lifetime of the electron-hole pairs, thicker cells may suffer
from larger recombination  rate and reduced efficiency.

The question is thus how to increase the optical path lengths of near
band gap photons inside the silicon, without increasing wafer
thickness. Several approaches have been tried, including texturing of
the wafer front or rear surface in various patterns and on different length
scales. For length scales larger than the wave length, the
incoming light is refracted into angles more parallel to the
wafer.\cite{Green} For length scales close to or smaller than the wavelength,
diffraction may couple light into guided modes.\cite{Sheng} However,
the texturing will often lead to an increase in surface defect states
and thereby increase recombination rates.

As an alternative to texturing, it has been proposed to place
metallic nanoparticles near the surface of the wafer.\cite{Catchpole}
The nanoparticles scatter the incoming light through a surface plasmon
resonance. Surface plasmons, or surface plasmon polaritons,
are electron density fluctuations at the interface between a metal and
a dielectric material. For a good introduction, see, e.g.,
Raether\cite{Raether} for plasmons in general, and
Bohren\&Huffman\cite{Bohren} for plasmons on  small particles. On the
surface of nanoparticles, the plasmons can be excited by an incoming
plane wave, and they exhibit a marked, tunable resonance. For
frequencies near the resonance, nanoparticles have an optical cross
section much larger than their geometrical cross section.  If this
resonance could be tuned to match the band gap of silicon, near
bandgap photons could be absorbed into the plasmon state with high
probability, while higher energy photons would be unaffected. Certain
progress has already been made in the plasmon tuning, e.g., at the
University of New South Wales, \cite{Pillai1, Pillai2} but further
development is required.

We believe that the energy of the surface plasmons can then be used to
create electron-hole pairs in two ways. First, the energy can be
emitted as light in directions along the wafer. This gives a longer
optical path inside the wafer, and thereby increases the indirect
absorption.\cite{Pillai1}

Second, the near field of the nanoparticles can excite electron-hole
pairs without phonon assistance, the momentum being transferred to the
nanoparticle.  This second process has to our knowledge not been
considered in the literature, and is the subject of the present paper.
Our results indicate that this mechanism will give an extra
contribution to the electron-hole pair generation, compared to
estimates that only take into account the
re-radiation of power, increasing the relative benefit of introducing
the nanoparticles.

Some of the plasmon energy goes into heating of the nanoparticles and
is obviously lost. This loss should be compared with the losses due to
the limited optical path when not exploiting the plasmons, or, if
texturing is used to increase the optical path length, with increased
losses due to recombination at interfaces. The plasmons will give an
improved efficiency if the losses to heating are smaller than previous
losses due to optical path length or recombinations.
If the resonance is
properly tuned, the only photons significantly affected will be those
that would otherwise be lost. Any fraction of this near bandgap light
that can be used efficiently contributes to a net gain for the cell.

For particles larger than the wavelength, a large fraction of the
light will be reflected rather than excite plasmons.\cite{Bohren} We
therefore consider only particles smaller than the wavelength of the
incoming light.  For such particles, the plasmons can be approximated
by a dipole mode, corresponding to uniform polarization of the
nanoparticle.\cite{Bohren} While the dipole approximation is usually
only accepted for particles with diameter less than one tenth of a
wavelength, we accept it as a first approximation for our order of
magnitude estimates.  We are not aware of any studies of how the near
field is changed by an interface between the layer embeddig
nanoparticles and the active layer of the solar cell.  For simplicity,
we will therefore restrict the further discussion to the case of an
electric dipole located inside an infinite medium consisting of
silicon. The dipole is excited by an incoming plane wave.

The far-field energy radiated from the dipole represents the maximum
energy that can be absorbed by indirect absorption. In real
applications, some of this light will inevitably be lost. The presence
of an interface may also increase the total emission,\cite{Mertz,
Benisty} but for the sake of our order of magnitude estimates, we will
ignore this effect.

The goal of this paper is to demonstrate that the direct absorption
effect should be considered when modeling the effect of plasmons, and
that it may have important implications for the optimal sizing and
positioning of the plasmons. The fact that plasmons can lead to an
increase in efficiency has been experimentally verified.\cite{Pillai2,
Derkacs} We therefore focus on the \textit{relative} importance of the
two mechanisms that could contribute to the increase, and how this
could influence cell design considerations.

\section{Theory} \label{Theory}

We use classical electrodynamics to describe both the nanoparticles
and the fields. The interaction with the silicon is described by
perturbation theory, and we use the tight binding model and the
parabolic approximation of the band gap extrema for the wave function
of the silicon.

We consider the incoming light to have a frequency close to the band
gap of silicon. This corresponds to $\hbar \omega = 1.1$ eV, or
$\omega \approx 10^{15}$ s$^{-1}$. We then get for the wavelength of
this radiation $\lambda = 2 \pi c / \omega \approx 1$ $\mu$m , and
from $c=\omega/k_p$ we get the photon wavenumber $k_p\approx 6\cdot
10^6$~m$^{-1}$. 

The vector potential due to a dipole is given as:\cite{Lorrain}
\begin{eqnarray}
\mathbf{A} 
&=& \frac{i \omega \mu_0}{4 \pi r} e^{-i(k_p r-\omega t)}\mathbf{p}_0= i
\mathbf{A}_0 A_r\, , \\
\mathbf{A}_0 &=& \frac{\omega \mu_0 k_p p_0}{4 \pi}\mathbf{e_p},\;\;\;\nonumber
A_r=\frac{1}{k_p r} e^{-i(k_p r-\omega t)} 
\end{eqnarray}
where $\mathbf{p}_0$ is the dipole moment, and $p_0 =
|\mathbf{p}_0|$. $r$ is the distance from the dipole.  Using the
previous rough estimates for $\omega$ and $k_p$ we get that $A_0/p_0
\approx 10^{15}$ Js/C$^2$m$^2$, while $A_r$ is a dimensionless
function containing all spatial dependencies of $\mathbf{A}$. The
magnitude of $p_0$ will be addressed later, but is not necessary for
the following comparisons of different terms.  The scalar potential
can be cast in the form $\Phi_0 \Phi_r$ where
\begin{equation}
\Phi_0=\frac{k_p^2 p_0}{4 \pi \epsilon_0}, \;\;\;
\Phi_r=\frac{\cos \theta}{k_p r}(i + \frac{1}{k_p r})e^{-i(k_p r -
  \omega t)}\, , 
\end{equation}
$\theta$ is the angle from the dipole axis.
We can estimate $\Phi_0/p_0 \approx 3 \cdot 10^{23}$ J/C$^2$m,
while $\Phi_r$ is again dimensionless.

The Hamiltonian of the system is:\cite{LL}
\begin{eqnarray}
  H&=&\frac{(-i \hbar \nabla+ e \mathbf{A})^2}{2m}-e\Phi  \\ &=&
-\frac{\hbar^2 \nabla^2}{2m}
-\frac{i e\hbar ( \nabla\cdot\mathbf{A}+\mathbf{A}\cdot\nabla)}{2m}
+\frac{e^2 \mathbf{A}^2}{2m}
- e\Phi
\end{eqnarray}
where $e$ is the positive elementary charge.
The $\mathbf{A}^2$-term can safely be neglected. 
Using the Lorentz gauge, $\nabla \cdot \mathbf{A} +
c^2 \dot{\Phi} = 0$, where $c =(\epsilon_0 \mu_0)^{1/2}$, we rewrite
the interaction Hamiltonian as
\begin{eqnarray}
 H_{\text{int}}&=& A_r \mathbf{C}_a\cdot\nabla -  \Phi_r C_b \, ; \\
\mathbf{C_a}&=&2\mu_{\text{B}} \mathbf{A}_0\, , \quad 
C_b = \nonumber
\left(
1 + \frac{\hbar \omega}{2 m c^2}
\right) e \Phi_0 
\end{eqnarray} 
where $\mu_B$ is the Bohr magneton.
Since $mc^2 \gg \hbar \omega$, $C_b\approx e \Phi_0$.
 
For the wave function, we use the standard tight binding
approximation,\cite{Singh} 
writing the wave function as 
\begin{equation}
\Psi_\mathbf{k}(r) = 
e^{i{\mathbf{k}\cdot\mathbf{r}}}\frac{1}{\sqrt{N_a}}
\sum_n \sum_l b_{\mathbf{k}l} \Psi_{nl}(r) 
\end{equation}
where $\Psi_{nl}(r) = \Psi_l(r-\mathbf{R}_n)$ is the $l$-th orbital
corresponding to the atomic wave function centered at the $n$-th atom
located at $\mathbf{R_n}$. The parameters
$b_{\mathbf{k}l}$ can in principle be found for each point in
$k$-space.  For states at the valence band maximum, there seems to be
good agreement between theory and experiment. For the conduction band
minimum, the fitting parameters are still optimized either for the position in 
$k$-space or for the effective mass in different directions, depending on what
is considered the most important. Based on Klimeck \textit{et
al.}\cite{Klimeck,Martins} we still assume that the minimum can be described by
a combination of single electron $p$, $s$ and $s^*$ states, where $s^*$ is
an excited $s$-state.

The transition rate for each $\mathbf{k'},\mathbf{k}$ can then be
found using Fermi's golden rule,
\begin{equation}
W_{\mathbf{k'k}} = \frac{2 \pi}{\hbar}\left| \langle
  \mathbf{k'}|H_{\text{int}}|\mathbf{k}\rangle \right|^2  
\delta(E_\mathbf{k'} - E_\mathbf{k}-\hbar \omega)
\end{equation}
where $\mathbf{k'}$ denotes the final state, and $\mathbf{k}$ the
initial one.  In the
following calculations, we will assume that the final state is near
the conduction band minimum. There are six equivalent such minima, the
effect of this will be addressed later. The absorbed energy by direct pair
creation is given as $P_d=\sum_{\mathbf{k',k}} \hbar \omega
W_{\mathbf{k',k}}$. We calculate this
using the parabolic approximation that is valid close to the band
edges. Since we are interested in initial states close to the top
of the valence band and final states close to a minimum in the
conduction band one can assume that the interaction matrix element is
weakly dependent on $\mathbf{k}$ and $\mathbf{k}'$,  $W_{\mathbf{k}'\mathbf{k}}\approx W_{\mathbf{k_0}\mathbf{0}}$. 

Writing $\mathbf{k''}=\mathbf{k}'-\mathbf{k_0}$ we get
\begin{equation}
E_\mathbf{k'} = E_g + \frac{\hbar^2 \mathbf{k}''^2}{2 m_c}, E_\mathbf{k} = - \frac{\hbar^2 \mathbf{k}^2}{2 m_v},
\end{equation}
and using
\begin{equation} 
\sum_{\mathbf{k}}= V \int d^3\mathbf{k} \frac{1}{(2 \pi)^3}=\frac{4 \pi V}{(2 \pi)^3}\int k^2 \frac{dk}{dE}dE\\
\end{equation}
this gives
\begin{eqnarray}
&\sum_{k}&W_{\mathbf{k}'\mathbf{k}}\delta(E_f-E_i-\hbar \omega)\\
\nonumber &=&
W_{\mathbf{k_0}\mathbf{0}}\frac{2^5 \pi^2 V^2}{(2 \pi)^6\hbar^6}(m_cm_v)^{\frac{3}{2}}\times\\
&\nonumber&
\int_{E_g}^{\infty} dE_f \int_{0}^{-\infty} -dE_i
\sqrt{(E_f-E_g)(-E_i)}\delta(E_f-E_i-\hbar \omega)\\
&=&
W_{\mathbf{k_0}\mathbf{0}}\frac{V^2(m_cm_v)^{\frac{3}{2}}}{2^4 \pi^3\hbar^6}(\hbar\omega - E_g)^2
\end{eqnarray}
where $m_c,m_v$ are the effective masses of the valence and conduction
bands, respectively, and $E_g$ is the gap energy.

Note that the energy dependence of the absorption is the same as that
of indirect absorption, rather than that for direct absorption in direct
band gap semiconductors, for which it is proportional to $(\hbar\omega
- E_g)^{1/2}$. The reason for this is the spread of Fourier components
in the dipole field, which take the role of the spread in phonon wave
numbers in the case of indirect absorption.

\section{Calculations}

We term the power emitted as radiation $P_r$, the power lost to
heating $P_h$, and the power going into direct electron-hole pair
generation $P_d$.

After standard calculations we get the absorption:
\begin{widetext}
\begin{eqnarray} \label{Pd}
P_d &=& \nonumber
\frac{2(m_c m_v)^\frac{3}{2}\omega (\hbar\omega-E_g)^2}{\hbar^6 k_p^2}
\left| 
\sum_{l'l} b_{\mathbf{k}'l'}b_{\mathbf{k}l} \left\{ 
 \frac{\alpha_{\mathbf{k'k}}\left(\langle l'|\mathbf{C_a} \cdot \nabla
|l\rangle + i \mathbf{C_a} \cdot \mathbf{k}\right)}{
(|\mathbf{k}'-\mathbf{k}|^2 - k_p^2)} + \frac{i\beta_{\mathbf{k'k}}C_b
\langle l'|l\rangle }{|\mathbf{k}'-\mathbf{k}|} 
\right\} \right|^2\, ;
\\
\alpha_{\mathbf{k'k}}&=&
\cos(|\mathbf{k}'-\mathbf{k}| r_a)
+\frac{i k_p}{|\mathbf{k}'-\mathbf{k}|}
\sin(|\mathbf{k}'-\mathbf{k}| r_a)
\, , 
 \nonumber \\
\beta_{\mathbf{k'k}} &=& \left\{
\frac{k_p \cos(|\mathbf{k}'-\mathbf{k}| r_a)+ i |\mathbf{k}'-\mathbf{k}| \sin(|\mathbf{k}'-\mathbf{k}| r_a)}{|\mathbf{k}'-\mathbf{k}|^2-k_p^2}
+\frac{\sin(|\mathbf{k}'-\mathbf{k}| r_a)}{|\mathbf{k}'-\mathbf{k}| k_p r_a}
\right\} \, .  
\end{eqnarray}
\end{widetext}
Here $r_a$ is the radius of the grain.  $\langle l'|\mathcal{O}|l\rangle = \int \Psi_{l'}^* \mathcal{O} \Psi_l d^3r$ denotes integration over the atomic orbitals for the operator $\mathcal{O}$. It can be assumed that the
elements of the sum where $n'\neq n$ will only give small corrections.

Writing $\mathbf{p}_0\cdot\nabla = \sum_{i=x,y,z}\mathbf{e}_i p_i
\nabla_i$, and having an initial state that is a combination of
 $p$-states, only the matrix elements $\langle s| e_i
 p_i\nabla_i|p_i\rangle$ do not vanish.
They are expected to be of the order of $1/a$ where $a$ is the lattice
constant, $5.4 \cdot 10^{-10}$ m for crystalline silicon. Regarding the
contributions of the scalar field and Umklapp processes, only the
elements with $l=l'$ do not vanish.

While each of the $x,y,z$ give different contributions depending on
the orientation of the dipole, it should be noted that there exist six
equivalent minima in the conduction band. While the matrix element due
to one minimum will be non-isotropic, the sum over all six minima is
expected to be isotropic and equivalent to two minima with
$\mathbf{p}_0\parallel \mathbf{k}$.
We define $k_0 = |\mathbf{k'}-\mathbf{k}|$. At the minima we have $k_0 \approx 0.85 \cdot 2 \pi /a \approx 10^{10}\mbox{m}^{-1}$. 

We see that all terms in Eq. \ref{Pd} show oscillations with period
$1/|\mathbf{k'}-\mathbf{k}|$ with increasing nanoparticle radius. As
the nanoparticle diameter cannot be expected to be well defined on
this length scale (atomic radius), we will simply take the average
over one period. We believe this to be justified both from considering
the limited coherence length of the electrons, and from the fact that
any physical measurement would include a dispersion of particle sizes.
While the limit of $r_a \rightarrow 0$ is
mathematically well defined, it is not physically meaningful, as it
describes a nanoparticle with less than one atom.  
 
Interestingly, the scalar potential provides much larger contribution
than the vector potential.
Keeping only the largest terms, we can make an order of magnitude estimate,
\begin{equation}
P_d \approx \frac{(m_c m_v)^\frac{3}{2}\omega e^2 p_0^2}
{32 \pi^2 \hbar^6 k_0^4 \epsilon_0^2}
\Delta E^2 \left(k_p+r_a^{-1}\right)^2
\end{equation}
$P_d$, $P_r$ and $P_h$ are all proportional to $p_0^2$. We define the
damping coefficients $\gamma_d = P_d/p_0^2$, $\gamma_r=P_r/p_0^2$,
$\gamma_h=P_h/p_0^2$ and $\gamma = \gamma_d+\gamma_r+\gamma_h$.  We
use these coefficients when comparing the importance of the different
mechanisms. To find the total absorbed power, we also need to estimate
the dipole moment, $p_0$,
which is determined by the amplitude of the incident wave, $E_0$, and 
the polarizability of the nanoparticles, $\alpha$, as
$p_0=\alpha(\omega) E_0$. While the polarizability is in general
dependent on the particle volume and shape, at the plasmon resonance
it is determined by the damping only. This can be shown from equating the power
absorbed from a plane wave by an oscillating dipole, $E_0p_0\omega/2$,
 with the total emitted power, $\gamma p_0^2$, giving 
\begin{equation}
\alpha = \frac{\omega}{2 \gamma(\omega)}.
\end{equation}
Assuming that the incident light is absorbed by a layer of
nanoparticles with a 2D 
density $n$, we get
\begin{eqnarray}
\frac{P_d}{W} &\approx& \frac{n (m_c m_v)^{3/2}\omega(\hbar \omega 
  -E_g)^2 e^2 \alpha^2 }{16\pi^2\hbar^6 c 
  k_0^4\epsilon_0^3}\left(k_p+r_a^ {-1}\right)^2 \nonumber \\
&\propto& 
\frac{\gamma_d}{(\gamma_d+\gamma_r+\gamma_h)^2} \, . \label{PdW}
\end{eqnarray}
Here $W=\epsilon_0 E_0^2 c/2$ is the incident 
power per area.  
As long as $\gamma_d$ is small compared to $\gamma_r+\gamma_h$,
it will not significantly change the polarizability, but if it becomes
of the same order, the decrease in polarizability will be more
important than the increase in absorption. This can easily be remedied
by a higher density of nanoparticles, but if the density becomes very
high, interaction between neighboring particles will change both the
polarizability and the plasmon resonance frequency.

\section{Results}
We see that the direct absorption
is proportional to
$(\Delta E)^2 \equiv(\hbar  \omega -E_g)^2$, the excess  energy after bridging
the band gap, squared. This is the same energy dependence as found for
indirect absorption if the single phonon processes including emission or absorption
of a phonon are considered  separately.

The resonance of the nanoparticles has a certain width, so $\Delta E$
also has a spread. To get a feeling for the order of magnitude of
$P_d$ we define $x=\Delta E/\hbar \omega$ and express the results
through this. We are interested in frequencies where direct absorption
of a plane wave would be impossible, so an $x$ of close to one is
irrelevant. The direct absorption shows no explicit
dependence on temperature, as opposed to indirect absorption. This may
indicate  a method  for  differentiating between  direct and  indirect
absorption in solar cells containing nanoparticles.  The possibility of changes in the band structure of silicon with temperature should still be considered.

The direct absorption, $P_d$, 
should be compared with the energy lost to
heating, $P_h$ or by radiation, $P_r$.
From Eq.~(\ref{Pd}) we have
\begin{equation}
P_d = \gamma_d p_0^2,\quad \gamma_d\approx 3\cdot 10^{44}x^2\left(1+\frac{1}{(k_p
    r_a)} \right)^2\,\frac{\mbox{J}}{\mbox{sm}^2\mbox{C}^2}\, ,
\end{equation}
where $x$ will usually be significantly less than one. With $50$ nm
particle radius we get $k_p r_a\approx 0.3$.  The total integrated
radiation from a dipole with the same dipole moment, $p_0$ is:
\cite{Lorrain}
\begin{equation}
P_r = \frac{c}{12 \pi \epsilon}\frac{p_0^2 (2
  \pi)^4}{\lambda^4}=\gamma_r p_0^2 \, .
\end{equation}
Using that $k_p \lambda = 2 \pi$ we obtain
\begin{equation}
\gamma_r = \frac{c k_p^4}{12 \pi \epsilon}\approx 10^{44}\, 
\frac{\mbox{J}}{\mbox{sm}^2\mbox{C}^2}\, 
\end{equation}
giving
\begin{equation}
\frac{\gamma_d}{\gamma_r} \approx 3 x^2 \left(1+\frac{1}{(k_p
    r_a)^2} \right).
\end{equation}
As shown in figure 1, we see that for sufficiently small
particles and if $x$ is not too small, $\gamma_d$ can be of the same
magnitude as $\gamma_r$, or even larger.  However, the gain for small
particles requires that the particle is very close to the
silicon. Designs where the nanoparticles are located outside the
silicon may lose the benefit of the $1/k_p r_a$ term.

In the previously mentioned experiments, the nanoparticles were
located close to a thin silicon wafer. For dipoles located at such an
interface, the total radiation increases, and a large fraction of the
radiation is directed into guided modes in the silicon
wafer.\cite{Mertz, Benisty, Catchpole, Pillai1}  For the light in the
bound modes, it is assumed that the optical path length is sufficient to
allow most of the radiation to be absorbed in indirect electron hole
pair creations. At the same time the areas where the near field is
strongest, has no silicon to absorb the energy. Under such
circumstances, it should be assumed that $\gamma_r$ dominates
$\gamma_d$.

However, there is also a possibility that radiated energy can excite a
plasmon on a neighboring particle, then again to be reemitted. This
would reduce the positive contribution of the plasmons in
architectures where the radiation along the wafer is exploited, as
more energy would be lost to heat. If we instead place the nanoparticles
inside the silicon and exploit the direct absorption, less energy
would be reabsorbed by neighboring nanoparticles, and thereby less
would be lost to heat.

Using a simple resistivity argument, $\gamma_h$ can be estimated as
\begin{equation}
\gamma_h\approx \frac{3\omega^2 \rho}{4 \pi r_a^3}\approx
2\cdot 10^{42}\frac{1}{(k_p r_a)^3} \frac{\mbox{J}}{\mbox{sm}^2\mbox{C}^2}
\end{equation}
for silver. For small particles, heating will take over as the
dominant damping mechanism, as shown in figure 1.

\begin{figure}[hb]
\begin{center}
\includegraphics[width=3in]{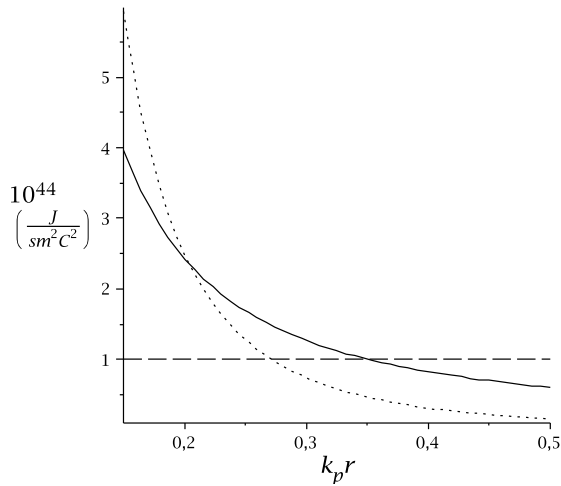}
\end{center}
\caption{The three damping mechanisms as function of $k_pr$. For small $k_pr$, $\gamma_h$ dominates(dotted line), for large $k_pr$, $\gamma_r$ dominates (dashed line), while $\gamma_d$ (solid line) may dominate in the middle region if $x$ is sufficiently large (here for $x=0.15$)}
\end{figure}

\section{Conclusions}
Our findings indicate that direct absorption due to surface plasmons
on metal nanoparticles does occur, and may give important corrections
to the total absorption for realistic parameters.

The direct absorption has been found to be
\begin{itemize}
\item proportional $\Delta E^2 = (\hbar \omega - E_g)^2$;
\item independent of temperature;
\item inversely proportional to the k-space position of the conduction
  band gap minimum to the fourth power;
\item comparable in magnitude to radiated energy in some cases.
\end{itemize}
The existence of the direct absorption mechanism is an argument for
placing the nanoparticles inside the silicon, rather than in front of,
or at the rear of the cell. This gives the additional requirement that
the problem of recombination centers at the particle surface can be kept to
a minimum.

Ideally, the size of the nanoparticles should be so small that
$\gamma_d$ dominates $\gamma_r$, but not so small that heating takes
over as the dominant mechanism. We assume that diameters from about a
tenth to half a wavelength could be suitable, depending on the
conductivity of the nanoparticle. The plasmon resonance should be
tuned using choice of material and particle shape (flattened for
red-shift\cite{Bohren}), to match the band gap of silicon.

The main questions that remain unanswered in our study concern the
effects of interfaces and surface states for the direct absorption.
We have not considered how an interface changes the near field, and
the presence of surface states in the silicon may significantly change
the problem in unpredictable ways. As the main contribution is from
very near the dipole, both surface electron states and the alteration
of the field due to an interface may be very important. There are also
some unaddressed problems related to the averaging over particle radii
and the finite coherence length of the electrons in the silicon.

It is possible that higher order modes will give larger contributions
to the absorption, these modes have been shown to be significant for
nanoparticles of sizes where reradiation is larger than heating.

The authors wish to acknowledge the financial support of the Norwegian
Research Council. We wish to
thank Alexander Ulyashin for introducing
us to the idea of using nanoparticle surface plasmons in solar cells,
and the SPREE at UNSW for giving us the necessary clues to get started. 

\end{document}